\pdfoutput=1

\documentclass[aps,twocolumn,showpacs,showkeys,superscriptaddress,amsmath,amssymb,nofootinbib,floatfix,prc]{revtex4-1}

\usepackage{subfigure}
\usepackage{bm}
\usepackage{hyperref}
\usepackage{graphicx,color}

\begin{document}

\hbadness=10000

\title{Cancellation of the sigma meson in thermal models}

\author{Wojciech Broniowski}
\email{Wojciech.Broniowski@ifj.edu.pl}
\affiliation{Institute of Physics, Jan Kochanowski University, PL-25406~Kielce, Poland}
\affiliation{The H. Niewodnicza\'nski Institute of Nuclear Physics, Polish Academy of Sciences, PL-31342 Krak\'ow, Poland}

\author{Francesco Giacosa}
\email{fgiacosa@ujk.edu.pl}
\affiliation{Institute of Physics, Jan Kochanowski University, PL-25406~Kielce, Poland}
\affiliation{Institute of Theoretical Physics, Goethe University, 
D-60438 Frankfurt am Main, Germany}

\author{Viktor Begun}
\email{viktor.begun@gmail.com}
 \affiliation{Institute of Physics, Jan Kochanowski University, PL-25406~Kielce, Poland}

\begin{abstract}
The by now well-established
scalar-isoscalar resonance $f_{0}(500)$ (the $\sigma$ meson) seems potentially relevant
in the evaluation of thermodynamic quantities of a hadronic gas, since its mass is low.
However, we recall that its contribution to isospin-averaged observables is, to a surprising accuracy, canceled by the repulsion from the
pion-pion scalar-isotensor channel. As a result, in practice one should not incorporate $f_0(500)$ in standard hadronic resonance-gas models
for studies of isospin averaged quantities.
In our analysis we use the formalism of the virial expansion,
which allows one to calculate the thermal properties of an interacting hadron gas in terms of
derivatives of the scattering phase shifts, hence in a model-independent way directly from
experimentally accessible quantities.
A similar cancellation mechanism occurs for the scalar kaonic interactions between the
$I=1/2$ channel (containing the alleged $K_{0}^{\ast}(800)$ or the $\kappa$ meson) and the $I=3/2$ channel.
\end{abstract}



\maketitle

\section{Introduction \label{intro}}

Nowadays, there is no doubt about the existence of the scalar-isoscalar
resonance $f_{0}(500)$ (traditionally known as the $\sigma$ meson). This very wide resonance
is best described through its pole position, which according to the Particle Data Group (PDG) lies in
a rather conservative range of $M_\sigma-i \Gamma_\sigma/2= \left(  400-550\right)  - i(200$-$350$)~\cite{Agashe:2014kda}.
Such a low range for the mass was first
reported in the 2012 version of the PDG tables~\cite{Beringer:1900zz}, while previous editions
quoted a very broad range $400$-$1200$~MeV. Detailed studies of this resonance
led to much smaller uncertainties: in Ref.~\cite{Caprini:2005zr} the pole at
$\left(400\pm6_{-13}^{+31}\right)  -i(278\pm6_{-43}^{+34}$) is obtained, while
in Ref.~\cite{Kaminski:2006qe,Kaminski:2006yv,GarciaMartin:2011jx,GarciaMartin:2011cn} the
value $(457_{-13}^{+14})-i(279_{-7}^{+11})$ is quoted (for a compilation of all estimates
and numerous references, see the mini-review
`Note on scalar mesons below 2~GeV' in Ref.~\cite{Agashe:2014kda}).

The question that we clarify in this paper is the role of the
resonance $f_{0}(500)$ in thermal models for relativistic
heavy-ion collisions (for reviews see, e.g.,
Refs.~\cite{Florkowski:2010zz,PBM:2003fa} and references therein).
Naively, one is tempted to include it as a usual but broad
Breit-Wigner resonance, as was done for instance in
Refs.~\cite{Torrieri:2004zz,Andronic:2008gu}. Notice that the pole mass of
$f_{0}(500)$ is light (even lighter than the
kaon), hence it might have a non-negligible effect on observables
in more accurate studies. Indeed, in Ref.~\cite{Andronic:2008gu}
the effect of $f_{0}(500)$ was found to increase the pion yield by
3.5\%.

However, broad resonances should be treated with care. Moreover, there are also repulsive channels
in the hadronic gas which must be taken into account on equal footing. The appropriate framework
here is the virial expansion (see, e.g., Ref.~\cite{Landau}). There, one uses only stable particles (pions, kaons, nucleons, ...) as
degrees of freedom, and their interactions are systematically incorporated in thermodynamics via coefficients of the
virial expansion. The second term of this expansion, corresponding to the $2 \to 2$ reactions, is straightforward to include, as it involves
the phase shifts accurately known from experiment.
Within this approach, the attraction due to the pole of $f_{0}(500)$ is
encoded in the pion-pion phase shift in the isoscalar-scalar $(I=J=0)$ channel.
In principle, this attraction could generate non-negligible
effects to thermodynamic properties. However, there is also
repulsion from the isotensor-scalar $(I=0,J=2)$ channel. It turns
out that this repulsion, whose effect is by construction not
included in standard thermal models, generates a negative
contribution to the second virial coefficient.  As a result, for
isospin-averaged observables where the isotensor channel has
degeneracy $(2I+1)=5$, one finds an almost exact cancellation of the
positive contribution due to $f_{0}(500)$ from the isotensor channel.
This feature has been reported already in Ref.~\cite{Venugopalan:1992hy}, and more recently
in Ref.~\cite{GomezNicola:2012uc}, where similar conclusions were reached
concerning the four-quark condensates and  scalar susceptibilities of the hadron gas.

As a net result of the cancellation, the
combined effect of $f_{0}(500)$ and the isotensor-scalar
channel  on all isospin-averaged properties of the hadron gas is
negligible. This is the basic result discussed in our study.

A similar, albeit partial, cancellation occurs also in the
pion-kaon $S$-wave interaction. The attraction in the channel $I=1/2$ is canceled
by a repulsion occurring in the $I=3/2$ channel. In the attractive $I=1/2$ channel many
studies find a kaon-like resonance, called a $K_{0}^{\ast}(800)$ meson
or $\kappa$~\cite{Lohse:1990ya,Dobado:1992ha}.

The framework of the virial expansion in the context of hot hadronic gas was used in numerous works, including
the nature of the repulsive terms~\cite{Welke:1990za,Venugopalan:1992hy,Eletsky:1993hv,Kostyuk:2000nx,Begun:2012rf,Arriola:2014bfa,Albright:2014gva},
meson properties~\cite{Dobado:2002xf}, thermal prediction from
chiral perturbation theory~\cite{Pelaez:2002xf}, or the treatment of the $\Delta(1232)$
resonance~\cite{Denisenko:1986bb,Weinhold:1996ts,Weinhold:1997ig}.

The paper is organized as follows: in Sec.~\ref{sec:pion_gas} we
briefly present the formalism and explicitly show the cancellation
between the isoscalar and the isotensor channels in the $S$-wave
(and also the $D$-wave) channels. In Sec.~\ref{sec:cancel} we
illustrate the cancellation with the trace of the energy-momentum tensor as a function of
$T$, as well as for the abundance of pions. In
Sec.~\ref{sec:kaons} we describe the similar case of the kaonic
resonance $\kappa$. Section~\ref{sec:correlations} contains a
discussion of two-particle correlation, where the cancellation
mechanism does not occur and where $f_0(500)$ plays a relevant
role~\cite{Broniowski:2003ax}. Finally, in
Sec.~\ref{sec:conclusions} we present further discussion and
conclusions.

\section{Interacting pion gas \label{sec:pion_gas}}

\subsection{The formalism}

In modeling relativistic heavy-ion collisions, the study of hadron
emission at freeze-out is successfully performed within thermal
models (for reviews see, e.g.,
Refs.~\cite{Florkowski:2010zz,PBM:2003fa} and references therein).
In this framework, the free energy of the hadron gas at
temperature $T$, i.e., $-k_B T \ln Z$, is represented as the sum
of contributions of all (stable and resonance) hadrons, whence
\begin{eqnarray}
\ln Z&=&\sum\limits_k \ln Z^{\rm stable}_{k}+\sum_k \ln Z^{\rm res}_{k},  \label{gas0}
\end{eqnarray}
where $k$ indicates the particle species.
In practice, one uses the list of existing particles from the PDG~\cite{Agashe:2014kda},

In the limit where the decay widths of resonances are neglected,
one has (for simplicity of notation we do not include chemical
potentials)
\begin{eqnarray}
\ln Z^{\rm stable, res}_{k}&=&f_{k} V \int \!
\frac{d^{3}p}{(2\pi)^{3}}\ln\left[  1\pm e^{-E_p/T}\right]^{\pm
1}, \label{gas1}
\end{eqnarray}
where $f_{k}$ is the spin-isospin degeneracy factor, $V$ is the
volume, $\vec{p}$ is the momentum of the particle, the mass of the
resonance is denoted as $M_{k}$, the energy is
$E_p=\sqrt{\vec{p}^2+M_k^2}$, and finally, the $\pm$ sign corresponds
to fermions or bosons, respectively. Once $\ln Z$ is determined,
all other thermodynamic quantities follow.

As a better approximation for the partition function, one can take
into account the finite widths of resonances. The quantity $\ln
Z_{k}$ is replaced by the integral over the distribution function
$d_{k}(M)$:
\begin{equation}
\ln Z^{\rm res}_{k}=f_{k}V \!\!\!
\int_{0}^{\infty}\!\!\!\!\!\!d_{k}(M)\, dM\int \!
\frac{d^{3}p}{(2\pi )^{3}}\ln\left[  1-e^{-E_p/T}\right]^{-1}.
\label{gas2}%
\end{equation}
For narrow resonances one can approximate $d_{k}(M)$ with a
(non-relativistic or relativistic) normalized Breit-Wigner
function peaked at $M_{k}$.
In the zero-width limit $d_{k}(M)=\delta(M-M_{k})$ and Eq.~(\ref{gas2}) reduces to Eq.~(\ref{gas1}).

For the case where the resonance is broad, or when we need to include interactions between stable particles which do not lead to resonances at all
(such as repulsion), we should use the virial expansion. Then the $2 \to 2$ reactions are incorporated according to the
formalism of Dashen, Ma, Bernstein, and
Rajaraman~\cite{Dashen:1969ep,Dashen:1974jw}. The
method, based on the virial expansion and valid for sufficiently low temperatures ($T \ll M$, where $M$ is the invariant mass of the pair),
uses as key ingredients the physical phase shifts which are well known for pion-pion scattering.
Then
\begin{eqnarray}
d_k(M)= \frac{d \delta_k(M)}{\pi dM}. \label{eq:dense}
\end{eqnarray}

For simplicity of notation we focus on the case where the only degrees of freedom are pions.
The partition function of the system up to second order in the
virial expansion is given via the sum
\begin{equation}
\ln Z=\ln Z_{\pi}+f_{IJ}\! \int_{0}^{\infty} \!\!\!\! dM \frac{d\delta_{IJ}}{\pi dM} \int \! \frac{d^{3}p}{(2\pi)^{3}}
 \ln\left[  1-e^{-E_p/T}\right]^{-1}, \label{lnzij}%
\end{equation}
where $\ln Z_\pi$ is the contribution from free pions evaluated according to Eq.~(\ref{gas1}),
$f_{IJ}=(2I+1)(2J+1)$ is the spin-isospin degeneracy factor.
In the case in which a series of narrow resonances labeled with
$k$ is present in the $(I,J)$ channel, the derivative of the phase
shift sharply peaks at the resonance positions according to
\begin{equation}
\frac{d\delta_{IJ}}{\pi dM}\simeq \sum_{k}
\frac{\Gamma_{IJ,k}}{2\pi}\left[  (M-M_{IJ,k})^{2}+\frac{\Gamma_{IJ,k}^{2}}{4}\right]^{-1},
\end{equation}
thus one recovers the contribution of the $(I,J)$ channel to the formula of
Eq.~(\ref{gas2}) in the Breit-Wigner limit. Moreover, for $\Gamma_{IJ,k}\rightarrow 0$ one
obtains  $d \delta_{IJ}/(\pi dM) = \sum_{k} \delta(M-M_{IJ,k})$,
in agreement with Eq.~(\ref{gas1})~\cite{Dashen:1974jw}.

Quantum-mechanically, the quantity $d\delta/(\pi dM)$ has a very simple interpretation: it is the difference of the density of
two-particle scattering states of the interacting and free systems. For completeness, we present the derivation of this well-known fact
in the Appendix.

\subsection{The cancellation}

In Fig.~\ref{fig:phase} we report the experimental $\pi\pi$ phase shifts as functions of the
invariant mass up to $1$~GeV and for different values of $(I,J)$, as parametrized in
Refs.~\cite{GarciaMartin:2011jx,GarciaMartin:2011cn,Nebreda:2011di}. From panel (a) it is clearly visible that the
channel $(0,0)$ is attractive (it is responsible for the emergence of the $f_{0}(500)$ pole),
while the channel $(2,0)$ is repulsive. The channel $(1,1)$
is also attractive and corresponds to the prominent $\rho$ meson. The isoscalar
and isovector $D$-wave channels ($J=2$) are reported in Fig.~\ref{fig:phase}(b), showing that the interaction strength in these channels is negligible.

\begin{figure}[tb]
\begin{center}
\includegraphics[width=.49\textwidth]{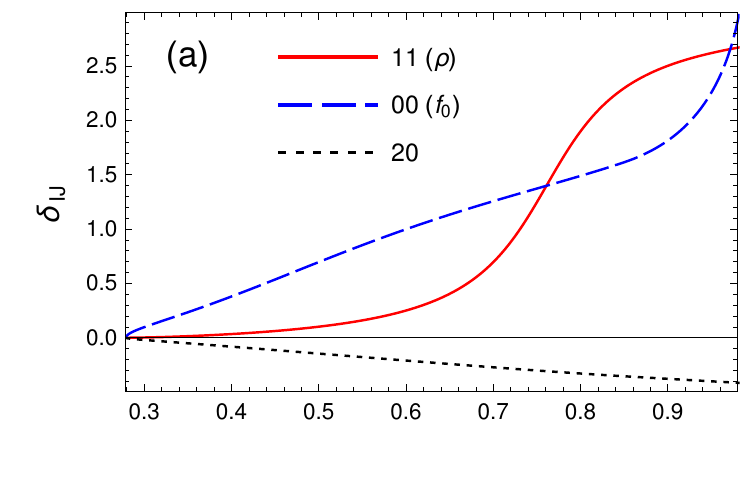} \\[0pt]\vspace{-9mm}
\includegraphics[width=.49\textwidth]{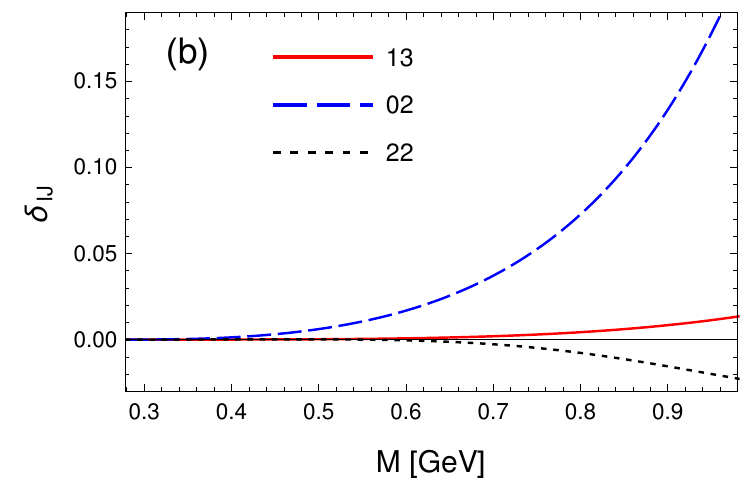}
\end{center}
\vspace{-7mm}
\caption{(color online) The parametrization of the pion-pion scattering phase shifts plotted vs invariant mass for various isospin-spin $(IJ)$
channels, taken from Ref.~\cite{GarciaMartin:2011jx,GarciaMartin:2011cn,Nebreda:2011di}.}%
\label{fig:phase}%
\end{figure}

\begin{figure}[tb]
\begin{center}
\includegraphics[width=.49\textwidth]{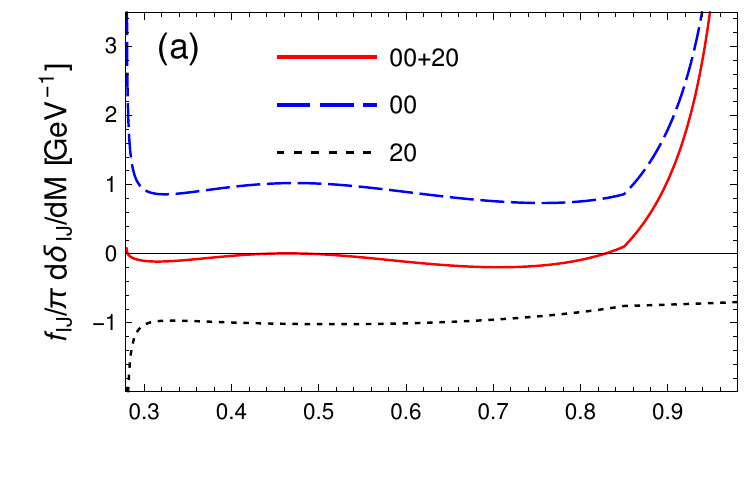}\\ \vspace{-9mm}
\includegraphics[width=.49\textwidth]{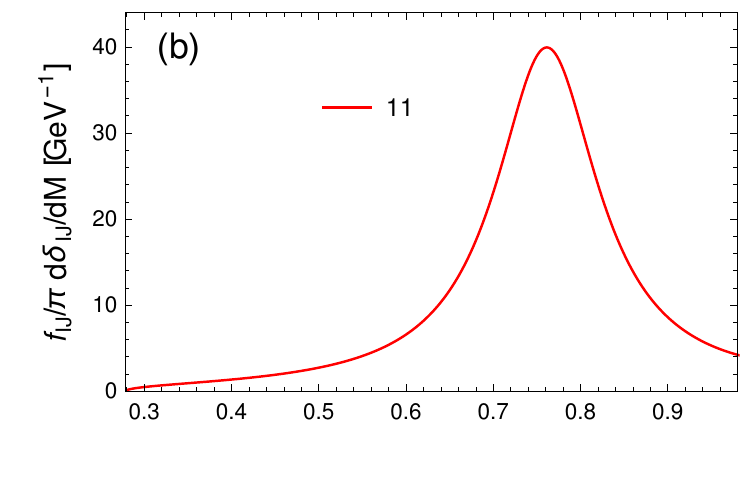}
\end{center}
\vspace{-7mm} \caption{(color online) The distributions $1/\pi
f_{IJ}\,d\delta_{IJ}/dM$, corresponding to the phase shifts of
Fig.~\ref{fig:phase}, plotted as functions of the $\pi\pi$
invariant mass $M$. We note a nearly complete cancellation for the
S-wave
interactions between the $I=0$ and $I=2$ channels up to $M\sim0.85$~GeV.
\label{fig:dphase}}
\end{figure}

In Fig.~\ref{fig:dphase} we present the derivatives of the phase
shifts from Fig.~\ref{fig:phase}, multiplied with $f_{IJ}/\pi$. We
note from panel~(a) that the resonance $f_{0}(500)$ does not lead
to a pronounced peak, but rather to a smooth plateau with an
integrable singularity close to the pion-pion threshold, which
follows from kinematics. Importantly, the isotensor-scalar channel
has a distribution, which up to $M \sim 0.85$~GeV is nearly a
mirror reflection of the $f_0$ channel. Note that this is achieved
with the multiplication of the isotensor channel by the isospin
degeneracy factor $(2I+1)=5$, which occurs for isospin-averaged
quantities (see Sec.~\ref{sec:correlations} for other cases). As a
result, the sum of the $(0,0)$ and $(2,0)$ channels nearly
vanishes until $f_{0}(980)$ takes over above $M \sim 0.85$~GeV. We
note that the negative phase shifts in certain channels may be
attributed to repulsion due to the finite size of hadrons, see,
e.g.~\cite{Welke:1990za,Venugopalan:1992hy,Eletsky:1993hv,Kostyuk:2000nx,Begun:2012rf,Arriola:2014bfa,Albright:2014gva}.

Recall that chiral perturbation theory ($\chi$PT) predicts
cancellation between the $S$-wave isoscalar and isotensor channels (for a historical review see, e.g., Ref.~\cite{Gasser:2009zz}).
The $S$-wave
scattering lengths $a_{I0}$, related to the phase shifts near the threshold via
$\delta_{I0}=q\,a_{I0}+o(q^2)$, with $q=\sqrt{{M^2}/{4}-m_{\pi}^2}$,
read explicitly at lowest order $\chi$PT~\cite{PhysRevLett.17.616}:
\begin{eqnarray}
a_{00}=\frac{7}{32\pi}\frac{m_{\pi}}{f_{\pi}^{2}},\;\;a_{20}=-\frac
{1}{16\pi}\frac{m_{\pi}}{f_{\pi}^{2}}.
\end{eqnarray}
We note the opposite signs, hence cancellation, while the relative
strength between the isoscalar and isotensor channels is $-3.5$.
Inclusion of chiral corrections~\cite{Colangelo:2000jc} brings the ratio to the value
$-4.95\pm 0.16$, while a global analysis including dispersion relations made in Ref.~ \cite{Pelaez:2004vs}
gives $-4.79\pm 0.55$. These values are compatible with $-5$ within the uncertainties.
While $\chi$PT explains the cancellation near the threshold, its occurrence until as
far as $M\sim0.85$~GeV is a quite remarkable fact which goes
beyond standard $\chi$PT.

In  Fig.~\ref{fig:dphase}(b) we display the $\rho$-meson channel, which dominates the
dynamics due to the fact that it is a pronounced resonance, and also carries a large spin-isospin
degeneracy factor.
In the remaining channels the interaction is very small. In addition, in the $D$-wave
channel there is also a partial cancellation between the $(0,2)$ and
$(2,2)$ channels at sufficiently low $M$.

\section{Cancellation of $f_{0}(500)$ in thermal models \label{sec:cancel}}

As the cancellation between the attraction from $f_0(500)$ and the repulsion from the isotensor-scalar
channel occurs at the level of the distribution functions $d_k(M)$, it will generically persist in all isospin-averaged
observables. The purpose of the examples in this section is to show, how much error one would make
by solely including the $f_0(500)$ and neglecting the isotensor repulsion.

\subsection{Trace of the energy-momentum tensor}

\begin{figure}[tb]
\begin{center}
\includegraphics[width=.49\textwidth]{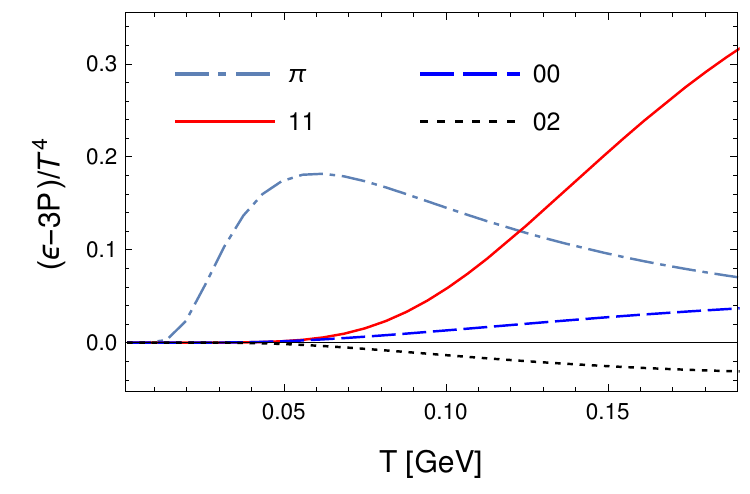}
\end{center}
\vspace{-7mm} \caption{(color online) Pion, $\rho$-meson,
isoscalar-scalar, and isoscalar-tensor contributions to the volume density of the trace
of the energy-momentum tensor divided by $T^4$, plotted as functions of $T$.
\label{fig:trace}}
\end{figure}

In Fig.~\ref{fig:trace} we show the contribution to the trace of the energy-momentum tensor of the hadron gas from various channels: as
expected, the contribution of a free pion gas is dominant at low $T$, while
the contribution of the $I=J=1$ channel ($\rho$ meson) rises fast and
overcomes that of free pions due to a large degeneracy factor. We note that the contribution from
the attractive isoscalar-scalar channel (dashed line) is nearly canceled by the repulsive isotensor-scalar channel (dotted line).

\subsection{Abundance of pions}

\begin{figure}[tb]
\begin{center}
\hfill \includegraphics[width=.48\textwidth]{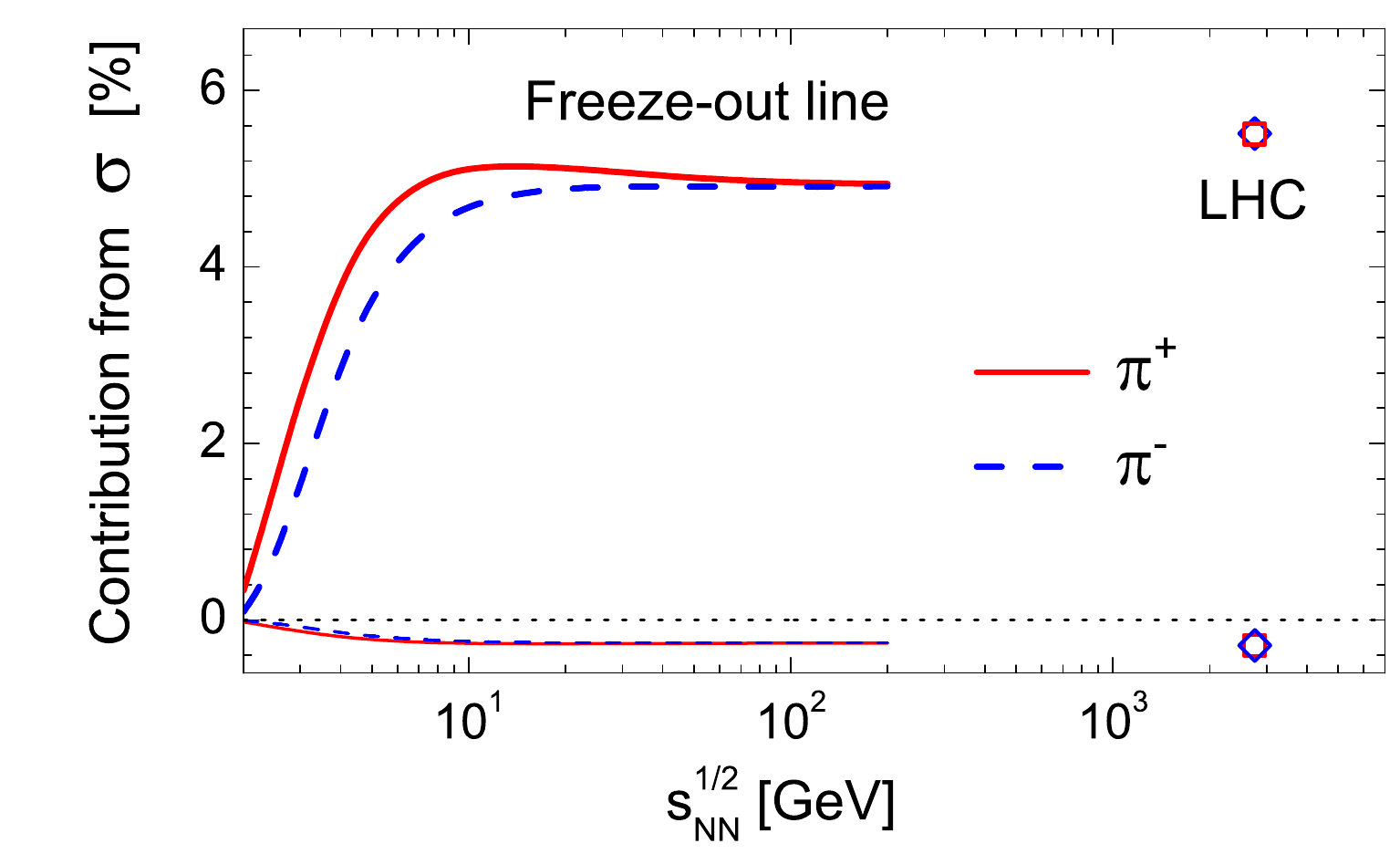}
\end{center}
\vspace{-7mm}
\caption{(color online) Relative contribution to the pion yield from $f_0(500)$, evaluated with SHARE~\cite{Torrieri:2004zz,Petran:2013dva}
applied to A+A collisions at various collision energies. The thick lines correspond to the naive Breit-Wigner implementation of $f_0(500)$, while
the thin lines indicate the correct inclusion of combined isoscalar-scalar and isoscalar-tensor channels. The points indicate the corresponding contributions at the LHC.}
\label{fig:RKpi}%
\end{figure}

The results of this subsection are of relevance to thermal modeling of particle production in ultra-relativistic
heavy-ion collisions. In a simple version of this approach, hadrons (stable and resonances) are assumed to achieve
thermal equilibrium, and later the resonances decay, feeding the observed yields of stable particles~\cite{Stachel:2013zma,Floris:2014pta}.
The basic outcome are thus the abundances of various hadron species, or their ratios, which allow to
determine the values of the thermal parameters at freeze-out. The contribution of light resonances is particularly important,
therefore the treatment of $f_0(500)$ is relevant.

In our study we use the SHARE~\cite{Torrieri:2004zz,Petran:2013dva} code  to investigate the effect of $f_{0}(500)$
on pion abundances. The purpose of this study is to see, how much pions are generated from the naive approach, where
$f_0(500)$ is improperly included as a wide Breit-Wigner pole with
$M_{\sigma}=484$~MeV and $\Gamma_{\sigma}/2=255$~MeV.
In Fig.~\ref{fig:RKpi} we show the thermal model
calculations carried out along the same freeze-out line as in~\cite{Begun:2012rf}, and for the temperature of $T=156$~MeV and
all chemical potentials equal to zero at the LHC~\cite{Stachel:2013zma,Floris:2014pta} (the extra LHC points in  Fig.~\ref{fig:RKpi}).
We find that with the naive implementation, the relative
feeding of the $f_0(500)$ to the pion abundances (i.e., the ratio of the number of pions originating
from the $f_0(500)$ to all pions) would be at a
level of up to 5\%, which is of a noticeable size and in agreement with the conclusions of Ref.~\cite{Andronic:2008gu}.

However, in reality this contribution is not there due to the
cancellation mechanism with the isotensor-scalar channel. The
proper implementation of both the scalar-isoscalar and
scalar-isotensor channels according to the phase-shift formula
(\ref{eq:dense}) for the density of states yields a very small
negative contribution of the combined scalar-isoscalar and
scalar-isotensor channels (at the level of $-0.3\%$) to the pion
yields, as indicated in Fig.~\ref{fig:RKpi} by the thin lines
and the point below zero for the LHC.

We note that the proper implementation described above leads to
lower pion yields, thus higher model hadron to pion multiplicity
ratios, at a level of a few percent. Of particular interest here are the  kaon to pion ratio, $K^+/\pi^+$, and the
proton to pion ratio, $p/\pi^+$.
For the former, the cancellation mechanism results in higher
values of the thermal model results. This may help to describe  the horn structure~\cite{Gazdzicki:1998vd} at
$\sqrt{s_{NN}}=7.6$~GeV~\cite{Alt:2007aa}, where models fall somewhat below the data~\cite{Andronic:2008gu}.
The latter is related to the LHC
proton-to-pion puzzle~\cite{Abelev:2012wca}, where the thermal model noticeably overpredicts  $p/\pi^+$. With the cancellation
of the $f_0(500)$ contribution, the proton-to-pion puzzle becomes even stronger,  opening
more space for possible novel interpretations~\cite{Begun:2013nga}.

\section{The case of $K_{0}^{\ast}(800)$ \label{sec:kaons}}

According to various works~\cite{Lohse:1990ya,Dobado:1992ha}, the attractive $\pi K$ channel
with $I=1/2$ and $J=0$ is capable of the generation of a pole
corresponding to the putative resonance $K_{0}^{\ast}(800)$. This resonance
has not been included in the summary table of the PDG, but it is naturally
expected to exist as the isodoublet partner of the established resonances
$f_{0}(500)$ or $f_{0}(980)$.
This wide resonance has a predicted mass of $\sim 680$~MeV and a very large decay width of
$\sim 550$ MeV, which makes its final assessment quite difficult (see also the
discussion in `Note on scalar mesons below 2~GeV' in Ref. \cite{Agashe:2014kda}).

\begin{figure}[tb]
\begin{center}
\includegraphics[width=.49\textwidth]{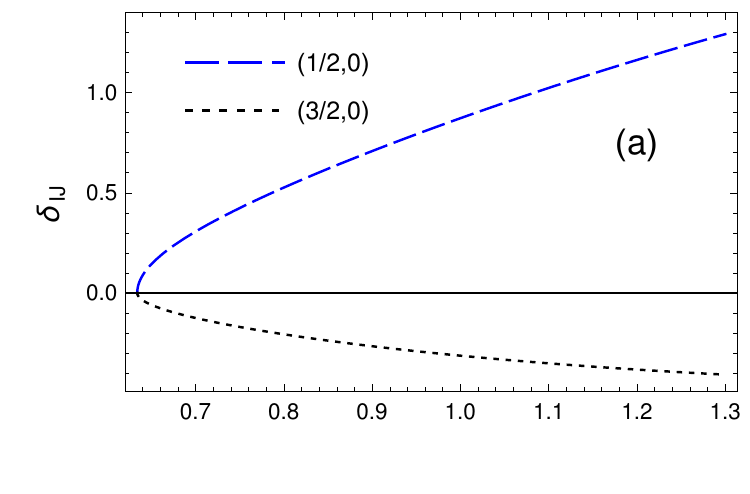}\\ \vspace{-9mm}
\includegraphics[width=.49\textwidth]{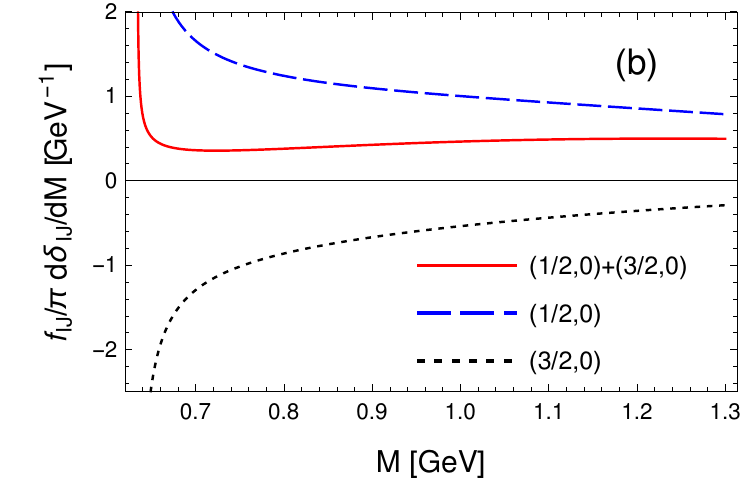}
\end{center}
\vspace{-7mm}
\caption{(color online) The pion-kaon S-wave phase shifts~(a) and their derivatives~(b), multiplied with the isospin-spin degeneracy factors, plotted as
functions of the invariant mass.
\label{fig:Kpi}}
\end{figure}

For the pion-kaon system, we use the effective-range fits to the
phase shifts given in Ref.~\cite{Estabrooks:1976sp}. The results
are plotted in Fig.~\ref{fig:Kpi}, where the phase shifts and
their derivatives (multiplied with the isospin-spin degeneracy
factors) for the channels $(0,1/2)$ and $(0,3/2)$ are presented.
Panel~(b) also displays the sum (solid line), which exhibits the
partial cancellation of the two contributions. Thus the conclusion
here is similar, but less pronounced, compared to the case of the
$\sigma$ meson.
The $I=1/2$ channel is dominated
by the vector channel $J=1$, represented by the resonance
$K^{\ast}(892)$.

\section{Correlations \label{sec:correlations}}

Up to now we have considered only isospin-averaged observables, where the
isospin degeneracy factor of \mbox{$(2I+1)$} is present, leading to cancellation between
the $(0,0)$ and $(2,0)$ channels. The cancellation does not occur for correlation variables.
Suppose, as an example, that we investigate the correlated $\pi^+ \pi^+$ pair production. Then, their source will be
the isotensor channel only, and none will come from $f_0$.

An investigation of this type has been reported in Ref.~\cite{Broniowski:2003ax}, devoted to the
analysis of the STAR Collaboration data~\cite{Adams:2003cc} on resonance production.
In particular, for the $\pi^+ \pi^-$
pairs, the isospin Clebsch factors yield for the number of pairs
\begin{eqnarray}
n_{\pi^+ \pi^-}(M) = 3 n_{\rho^0}(M) + \frac{2}{3} n_{f^0}(M) + \frac{1}{3} n_{(2,0)}(M),  \label{eq:res}
\end{eqnarray}
where the factor of 3 in front of the $\rho_0$ contribution comes from the spin degeneracy. Since up to $M\sim 0.85$~GeV (cf. Fig.~\ref{fig:dphase}a) the ratio
$n_{(2,0)}(M)/ n_{f_0}(M)\simeq -1/5$, the relative yield of the isotensor channel compared to the isoscalar channel
in Eq.~(\ref{eq:res}) is
only about 10\%, and there is no cancellation. We thus clearly see the potential importance of the $\sigma$ in studies of
pion correlations~\cite{Broniowski:2003ax}.

\section{Conclusions \label{sec:conclusions}}

In this work we have presented, in the framework of the virial
expansion, that the contribution of the resonance $f_{0}(500)$
(alias $\sigma$) to isospin-averaged observables (thermodynamic
functions, pion yields) is nearly perfectly canceled by the
repulsion from the isotensor-scalar channel. The cancellation
occurs from a ``conspiracy'' of the isospin degeneracy factor, and
the physical values for the derivative of the phase shifts with
respect to the invariant mass. Thus, the working strategy in
thermal models of hadron production in relativistic heavy-ion collisions, which incorporate stable hadrons and resonances, is to
leave out $f_0(500)$ from the sum over resonances. Including it,
would spuriously increase the pion yields at a level of a few
percent. Alternatively, one may use physical phase shifts to do
more accurate studies, but then, of course, all relevant isospin
channels must be incorporated. The validity of the conclusion is
for temperatures $T$ smaller than the invariant mass of the
hadronic pair. The lowest mass comes from the two-pion threshold,
hence $T < 300$~MeV, which is a comfortable bound.

A similar cancellation occurs for $K_{0}^{\ast}(800)$ (alias
$\kappa$) with quantum numbers $I=1/2, J=0$, whose contribution is
compensated by the $I=3/2, J=0$ channel. On the other hand, in
correlation studies of pion pair production, there is no
cancellation mechanism.

\begin{acknowledgments}
We are grateful to Robert Kami\'nski for providing the
state-of-the-art pion phase shifts used in our analysis. We also
thank Wojciech Florkowski, Jose Pelaez, Giuseppe Pagliara,
Valentina Mantovani Sarti, and Thomas Wolkanowski for useful
discussions. This research was supported by the Polish National
Science Center, Grant DEC-2012/06/A/ST2/00390.
\end{acknowledgments}

\appendix

\section{Derivation of the phase-shift formula \label{sec:phase}}

For completeness, in this Appendix we quote the quantum-mechanical
derivation of the phase-shift formula of Eq.~(\ref{gas2}).
The relative radial wave function of a pair of scattered particles
with angular momentum $l$, interacting with a central potential,
has the asymptotic behavior
$\psi_{l}(r)\propto\sin[kr-l\pi/2+\delta_{l}]$, where $k=|
\vec{k}|$ is the length of the three-momentum, and $\delta_{l}$ is
the phase shift. If we confine our system into a sphere of radius
$R$, the condition $kR-l\pi/2+\delta_{l}=n\pi$ with $n=0,1,2,...$
must be met, since $\psi_{l}(r)$ has to vanish at the boundary.
Analogously, in a free system $kR-l\pi/2=n_{\rm free}\pi$. In the
limit $R \to \infty$, upon subtraction,
\begin{eqnarray}
\frac{\delta_l}{\pi}=n - n_{\rm free}.
\end{eqnarray}
Differentiation with respect to $M$ yields immediately the
interpretation, that the distribution $d \delta_l/(\pi dM)$ is
equal to the difference of the density of states in $M$ of the
interacting and free systems~\cite{Landau}.


\bibliography{extras,fromhep}

\end{document}